\input{aipcheck}

\documentclass[
    ,final            
  ]
  {aipproc}

\usepackage{amsmath,amssymb,amsfonts,latexsym,cancel} 

\layoutstyle{6x9}

\begin{document}

\title{Rotochemical heating with a density-dependent superfluid energy gap in neutron stars}



\author{Nicol\'as Gonz\'alez-Jim\'enez}{
  address={Departamento de Astronom\'\i a y Astrof\'\i sica, Pontificia Universidad
        Cat\'olica de Chile, Casilla 306, Santiago 22, Chile.}
}

\author{Crist\'obal Petrovich}{
  address={Departamento de Astronom\'\i a y Astrof\'\i sica, Pontificia Universidad
        Cat\'olica de Chile, Casilla 306, Santiago 22, Chile.}
}

\author{Andreas Reisenegger}{
  address={Departamento de Astronom\'\i a y Astrof\'\i sica, Pontificia Universidad
        Cat\'olica de Chile, Casilla 306, Santiago 22, Chile.}
}

\begin{abstract}

When a rotating neutron star loses angular momentum, the reduction of the centrifugal force makes it contract. This perturbs each fluid element, raising the local pressure and originating deviations from beta equilibrium, inducing reactions that release heat (rotochemical heating). This effect has previously been studied by Fern\'andez and Reisenegger for neutron stars of non-superfluid matter and by Petrovich and Reisenegger for superfluid matter, finding that the system in both cases reaches a quasi-steady state, corresponding to a partial equilibration between compression, due to the loss of angular momentum, and reactions that try to restore the equilibrium.
However, Petrovich and Reisenegger assumes a constant value of the superfluid energy gap, whereas theoretical models predict density-dependent gap amplitudes, and therefore gaps that depend on the location in the star. In this work, we try to discriminate between several proposed gap models, comparing predicted surface temperatures to the value measured for the nearest millisecond pulsar, J0437-4715. 
\end{abstract}
\maketitle

	Rotochemical heating can keep neutron stars hot beyond their standard cooling time of $\sim10^{7}$ yr. As such a star spins down, the centrifugal force decreases and the star contracts. The density of any given fluid element increases, generating chemical imbalances. In the case of millisecond pulsars, these imbalances grow slowly, due to the steady spin-down, until about $\sim10^{6-7}$ yr, when beta decays begin to operate. At this point, the star reaches a quasi-steady state where the rate at which spin-down modifies the chemical equilibrium  is the same at which the reactions restore the equilibrium. In this state, one of the main results is that the temperature depends only on the current angular velocity and its time derivative \cite{fer05}.\\




The main effect of superfluidity is the suppression of beta reactions due to a gap at the Fermi surface that reduces the available phase space. \citet{petro} found that the inclusion of superfluidity raises the temperature of the quasi-steady state and delays the heating of the star. However, their model did not take into account the theoretical models of nucleon pairing, which predict a density-dependent amplitude of the energy gap \cite{anderson}. The implications of any given model are the total, partial, or no inhibitions of the beta reactions, depending on the location in the star.\\

	\begin{figure}[h]
\includegraphics[width=18cm]{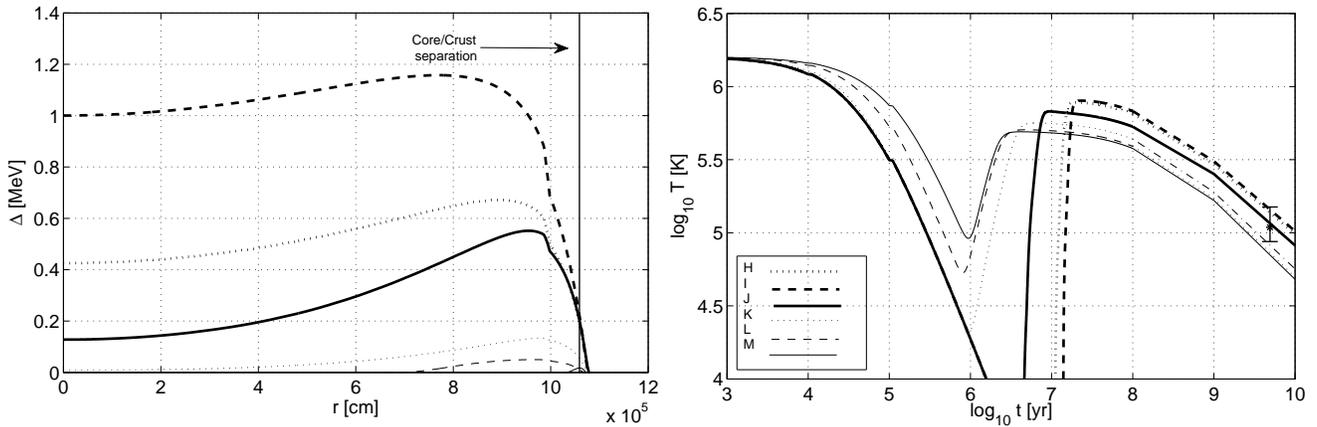}
\caption{{\bf Left panel}: Spatial distribution of the theoretical models obtained from \cite{anderson}.
{\bf Right panel}: Thermal evolution of a neutron star of $1.76\; M_\odot$, B= $10^{8}\; G$ and initial temperature $T_0= 10^8 \;K$, with the six superfluid models (assuming isotropic gaps). The error bar is the measured surface temperature of the nearest millisecond pulsar, J0437-4715 \cite{kar04}.}
\end{figure}

The left panel of Fig. 1 shows the dependence of the gap on the location in the star. Although the models predict an anisotropic gap, in our first approach we approximate the gap as being isotropic. The zone where the amplitude of the energy gap is smallest will dominate the reaction rates and thus the energy release. In the right panel, the thin lines are models whose spatial distribution in the core makes the behavior in the final stage of the thermal evolution practically the behavior of a non-superfluid star. The predicted temperatures for these models do not quite agree with the millisecond pulsar measurement \cite{kar04}. On the other hand, the models represented by the thick lines make the star completely superfluid. In this case, the reactions that heat the star will be forbidden until the chemical imbalances overcome the threshold given by the minimum amplitude of the energy gap in the core. Finally, when the reactions are allowed again, the star rapidly increases its temperature and reaches the quasi-steady state. Overall, all the models considered give fairly similar predictions, all of which are at least marginally consistent (within  2$\sigma$) with the only observation available so far \cite{kar04}.



\begin{theacknowledgments}
Research supported by FONDECYT Regular Project 1060644, FONDAP Center for Astrophysics (15010003), and Proyecto Basal PFB-06/2007. 

\end{theacknowledgments}



\bibliographystyle{aipproc}   


\IfFileExists{\jobname.bbl}{}
 {\typeout{}
  \typeout{******************************************}
  \typeout{** Please run "bibtex \jobname" to optain}
  \typeout{** the bibliography and then re-run LaTeX}
  \typeout{** twice to fix the references!}
  \typeout{******************************************}
  \typeout{}
 }

\end{document}